# The design and test results of A Giga-Bit Cable Receiver (GBCR) for the ATLAS Inner Tracker Pixel Detector

L. Zhang,[a,b] D. Gong[a,1] T. Liu,[a] C. Chen,[a] B. Deng,[c] S. Hou,[d] G. Huang,[b] X. Huang,[a] C. Liu,[a] P. Moreira,[e] Q. Sun,[f] X. Sun,[b] S. Kulis,[e] J. Ye,[a] and W. Zhang,[a]

[a] *Southern Methodist University,*
    *Dallas, TX 75275, U.S.A.*

[b] *Central China Normal University,*
    *Wuhan, Hubei, 430079, P.R. China*

[c] *Hubei Polytechnic University,*
    *Huangshi, Hubei, 435003, P.R. China*

[d] *Academia Sinica,*
    *Nangang, Taipei, 11529, Taiwan*

[e] *CERN,*
    *1211 Geneva 23, Switzerland*

[f] *Fermi National Accelerator Laboratory,*
    *Batavia, IL, 60510, U.S.A.*

E-mail: dgong@smu.edu

ABSTRACT: This paper presents the design and test results of a Gigabit Cable Receiver ASIC called GBCR for the HL-LHC upgrade of the ATLAS Inner Tracker (ITk) pixel detector. Three prototypes (GBCR1, GBCR2, and GBCR3) have been designed in the CERN-identified 65 nm CMOS technology. GBCR receives seven (GBCR2) or six (GBCR3) channels (RX) each at 1.28 Gbps from the front-end readout chip RD53B via flex cables up to 1 meter and Twinax cables up to 5 meters and sends the equalized and retimed signals to lpGBT. Both GBCR2 and GBCR3 ASICs have two transmitting channels (TX) that pre-emphasize the signals from lpGBT before sending them to RD53B through the same cables. No Single-Event Upset (SEU) is observed in any tested channels of GBCR2 in a 400 MeV proton beam. The extrapolated bit error rate for the future HL-LHC application is below $8\times10^{-16}$, significantly below the specified BER criterion. GBCR3 is designed to improve the immunity to single-event-upset by applying the Triple Modular Redundancy (TMR) technology to all RX channels. The retimed signals from GBCR3 have less total jitter than those from GBCR2 (35 ps versus 79 ps). Each receiver channel of GBCR3 consumes 75% more power than that of GBCR2.

---

[1] Corresponding author.

## 1. Introduction

As shown in figure 1, the Inner Tracker (ITk) [1] of the ATLAS detector for the High-Luminosity Large-Hadron Collider (HL-LHC) era is read out by the ASIC RD53B [2]. The output signal is transmitted for the first few meters through electric cables. The Opto-Box uses lpGBT [3] and VTRx+ [4] to aggregate 1.28 Gigabit per second (Gbps) electric signals to a 10.24 Gbps signal and send it off the ATLAS detector via optical fibers. This low-mass electric cable has a section of flexible printed circuit followed by a Twinax cable (AWG34) of 3 to 6 meters. This cable induces significant Inter-Symbol Interference (ISI) due to the high-frequency signal loss. For example, the signal loss at the Nyquist frequency of the 1.28 Gbps uplink data rate is about 10 dB. The signal arriving at lpGBT does not meet its input specifications. To overcome this problem, a Gigabit Cable Receiver (GBCR) [5, 6] is designed to restore the signal. In addition to the multiple equalizing channels (RX), each GBCR provides two pre-emphasized transmitting channels (TX) to reduce the jitter of the 160 Mbps command signals, from which RD53B recovers the clock. Three GBCR prototypes (GBCR1, GBCR2, and GBCR3) have been designed in the CERN-identified 65 nm CMOS technology. In this paper, we present the irradiation test results of GBCR2 and the design and preliminary test results of GBCR3.

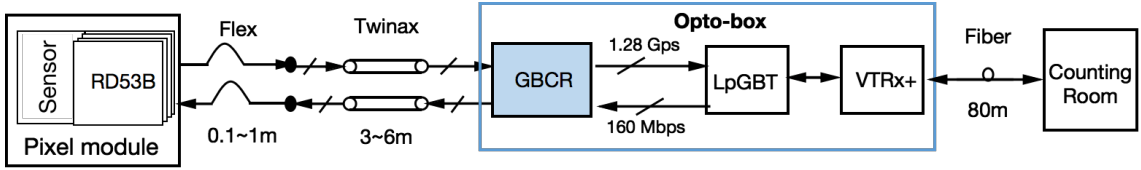

**Figure 1.** Block diagram of the ATLAS ITk readout system.

## 2. Design of GBCR2 and GBCR3

The block diagram of GBCR2 is shown in figure 2. GBCR2 has seven uplink receiver channels, two downlink transmitter channels, an $I^2C$ target module, and a phase shifter. Each uplink channel consists of an equalizer, a DC-offset-cancellation circuit, a limiting amplifier, a retiming unit, and a Current-Mode-Logic (CML) line driver. The clock of the retiming unit is from an on-chip phase shifter shared by all channels. GBCR2 operates in either an equalizer or retiming mode. In the equalizer mode, the phase shifter is turned off and the retiming unit is bypassed. The cable degenerated signal is equalized and amplified before it is sent to a differential output driver. In the retiming mode, the phase shifter is turned on. The equalized and amplified signal is re-sampled with a clock from the phase shifter to reduce the jitter. For each downlink channel, the signal from lpGBT is pre-emphasized and sent to RD53B through the same type of cables as the uplink channels. As for the downlink channel, a fine-tuning pre-emphasizer is implemented in a Continuous Time Linear Equalization (CTLE) resistor-capacitor network. The detailed design of GBCR2 can be seen in [5, 6].

GBCR3 is designed to improve the immunity to SEUs of the uplink channels and strengthen the pre-emphasis of the downlink channels. For uplink channels, we applied the Triple Modular Redundancy (TMR) technology. The block diagram of the TMR structure in the uplink channel of GBCR3 is shown in figure 3. By simulating the design with a charge injection at various nodes, we found that the equalizer and the passive attenuator in GBCR2 are most sensitive to potential SEU events. The passive attenuation, which was proved to be unnecessary, was removed in GBCR3. The TMR technology is applied to the equalizer, the limiting amplifier, and the DC-offset cancellation circuit. A majority voter is added after the limiting amplifier. A multiplexer selects one of the three unvoted signals (for test purposes), the voted signal, or the retimed signal to output. For downlink channels, the pre-emphasis strength is larger than 15 dB. GBCR3 has larger programmable resistance and capacitance than GBCR2.

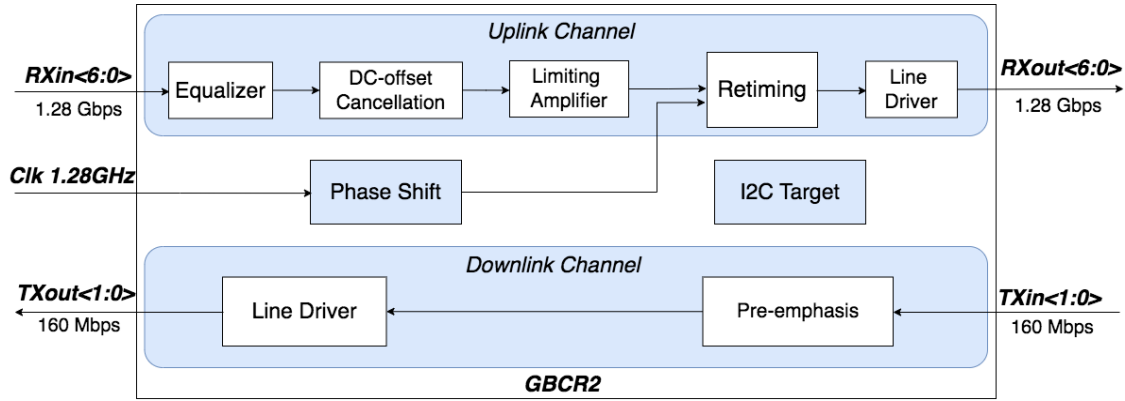

**Figure 2.** Block diagram of GBCR2

Both GBCR2 and GBCR3 operate at a single supply voltage of 1.2 V. The die is 1 mm × 3 mm. We reduce the number of uplink channels from 7 in GBCR2 to 6 in GBCR3, so that GBCR2 and GBCR3 have the same die size. GBCR is packaged in a 48-pin Quad-flat No-leads (QFN) package. The packaged chip is 6 mm × 6 mm. GBCR3 is pin-to-pin compatible with GBCR2, except that the last uplink channel in the GBCR3 package is not connected. GBCR3 can use the test board and the Opto-Box design for GBCR2. The main specifications of the two ASICs are summarized in Table 1.

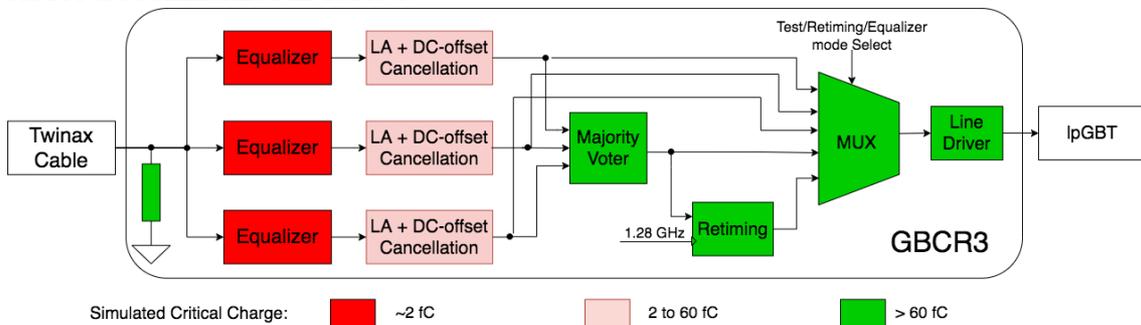

**Figure 3.** Block diagram of the TMR structure in the uplink channel.

**Table 1**. The main specifications of GBCR2 and GBCR3.

| Name | Number of Channels | Die size (mm × mm) | Supply Voltage (V) | Power consumption (mW) | Package | Total jitter (ps, Pk-Pk) |
|---|---|---|---|---|---|---|
| GBCR2 | Uplink: 7 Downlink: 2 | 1 x 3 | 1.2 | 250 | QFN48 | 200 |
| GBCR3 | Uplink: 6 Downlink: 2 | 1 x 3 | 1.2 | 350 | QFN48 | 200 |

## 3. Irradiation test results of GBCR2

The GBCR2 [6] has passed all functional tests in the lab environment. In the full-chain readout test, the uplink channels meet the requirements, but the downlink channels are marginal. In this section, we focus on the GBCR2 irradiation performance of single-event effects. The other performances of GBCR2 will be discussed together with those of GBCR3 in Section 4. GBCR is expected to be exposed to $2.33\times10^{14}$ hadrons/cm$^2$ with energy higher than 20 MeV in the HL-LHC operation time of 10 years (the active time of $1\times10^8$ seconds). The Bit Error Rate (BER) of each TX or RX channel is specified to be below $1\times10^{-12}$.

GBCR2 was irradiated at the Irradiation Test Area of Fermi National Accelerator Laboratory (FNAL) in Chicago, USA. The block diagram of the test setup is shown in figure 4(a). A GBCR2 chip was mounted 18 mm away from the edge of the carrier board. The carrier board was shielded with 16−µm-thick aluminum foil. The carrier board was inserted into a mechanical crate. The crate was installed on a 2-dimension stage so that the x and y positions of the chip relative to the beam (the z axis) were adjustable. During the test, the center of the beam was 67 mm or 30 mm away from the chip. At these distances, the board and the chip were shielded from the beam's Electro-Magnetic Interference (EMI). An FPGA evaluation board (Xilinx KC705) generated a Pseudo-Random Binary Sequence (PRBS) $2^7$-1 and sent it to the GBCR2 chip. The FPGA also checked the data that was looped back for errors. Daisy-chained Display Port cables were employed between the carrier board and the FPGA to mimic the flex+Twinax cables in the ATLAS application. A clock generator board (Silicon Labs Si5338EVB) provided a 160 MHz reference clock to the FPGA. The FPGA, the clock generator, and the power supply unit were placed about three meters away from the GBCR2 chip and shielded behind a concrete wall in the beam room. A Personal Computer (PC) configured the FPGA via a USB cable. The error log was transmitted from the FPGA to the PC via an Ethernet cable. The PC was in the counting room 30 meters away from the beam chamber. Due to a test setup limitation, the first transmitter channel (TX1) was not tested. To monitor the operation status of the test system, a bit error was injected on purpose into each channel every 26.8 seconds. This "heartbeat" was removed in the offline data analysis. Figure 4(b) shows a photo of the test setup. Figure 4(c) is a photo of the carried board shielded with aluminum foil in the crate. Figure 4(d) is a photo of the GBCR2 carrier board with the shielding aluminum foil. Figure 4(e) is a photo of the carried board of GBCR2.

GBCR2 was irradiated in a 400 MeV proton beam. In each minute, protons are delivered during a spill, whose length is set to be 7 µs. The beam profile is a 2-dimension Gaussian shape. A 1-mm-thick 25-mm-square aluminum sheet was placed on the back of the carrier board. The center of the sheet was aligned with the chip. The radioactivity of the isotopes $^{22}$Na and $^{7}$Be from the aluminum sheet was measured after the test. Based on the radioactivity of the sheet, we estimated the standard deviation of the beam profile to be about 20 mm.

No error was observed in any channel for about 2 hours at 67 mm and about 4 hours at 30 mm. Neither single-event latch-ups nor other SEEs were observed. The fluxes at 67 mm and 30 cm are estimated to be $1.7\times10^{13}$ p/cm$^2$/s and $1.4\times10^{15}$ p/cm$^2$/s, respectively. The total fluence of the test periods is $2.3\times10^{12}$ p/cm$^2$. The cross-section of the SEEs is estimated to be less than $4.3\times10^{-13}$ cm$^2$. The extrapolated BER in the future HL-LHC application [7] is below $8\times10^{-16}$, significantly less than the specified BER criterion.

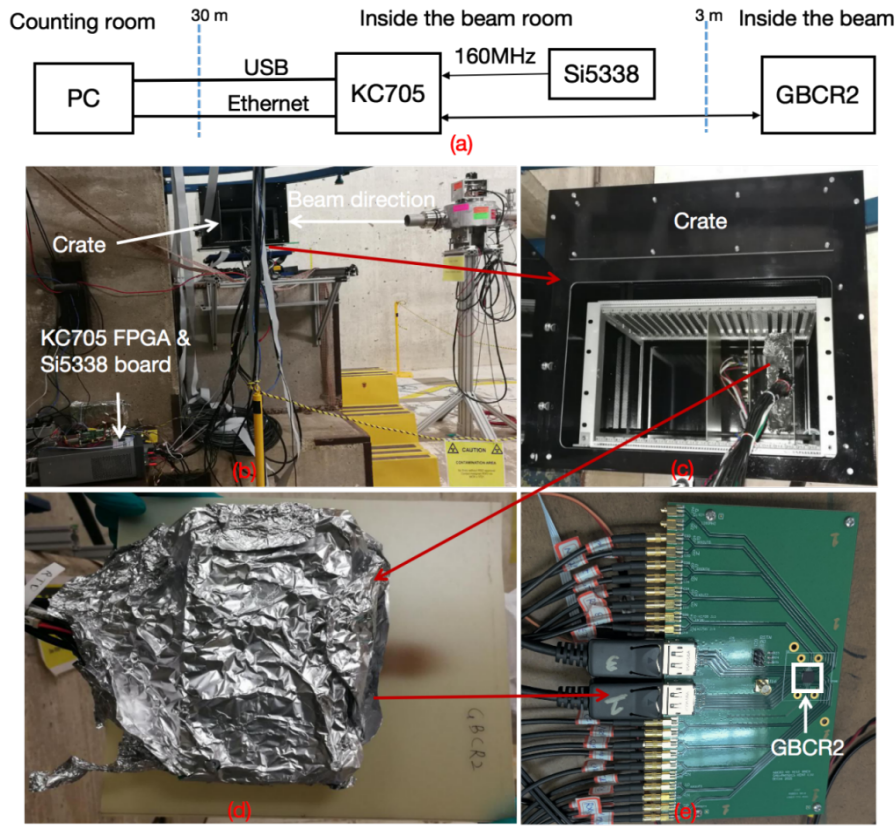

**Figure 4.** Block diagram of the irradiation test setup (a), photos of the irradiation test setup (b), a metal box in the beam box (c), the GBCR2 board with shielding aluminum foil (d), and the GBCR2 carried board (e).

The supply current of GBCR2 was monitored. The Total Ionizing Dose (TID) exposed in the tests is 2.3 kGy. No significant change in the supply current (less than 1%) was observed. This result of GBCR2 is consistent with that of GBCR1, which has the same major functional blocks as GBCR2. GBCR1 was irradiated in $^{60}$Co gamma rays to 200 kGy. No significant degradation was observed after the test.

## 4. Preliminary test results of GBCR3

GBCR3 was preliminarily tested in the lab. Typical eye diagrams of an uplink channel in the equalizer and retiming modes are shown in figure 5. The total jitters are 233 ps and 35 ps (peak-peak) in the equalizer and retiming modes, respectively. GBCR2 has lower total jitter (129 ps, peak-peak) in the equalizer mode, but higher total jitter (79 ps, peak-peak) in the retiming mode. The jitter increase in the equalizer mode of GBCR3 is attributed to the triplicated layout and the majority voter. The improvement in the total jitter in GBCR3 in the retiming mode benefits from the optimized layout of the retiming unit. The layout of the retiming circuit in GBCR3 is improved. The length of the differential clock traces in GBCR3 decreases, whereas the gaps between the neighbor traces increase. Besides the layout optimization, a high-power buffer is added to drive the retiming block and reduce the potential ISI jitter. The recovered signal jitters of both GBCR2 and GBCR3 in the retiming mode meet the lpGBT's input jitter requirement of less than 200 ps. The power consumption of the uplink channel is 22.1 mW and 22.5 mW in the equalizer and retiming modes, respectively. Due to the TMR used in the design, the power dissipation per receiver channel of GBCR3 is increased by 75% in the equalizer mode and 55% in the retiming mode. More performance tests of downlink channels and irradiation tests will be carried out in the future.

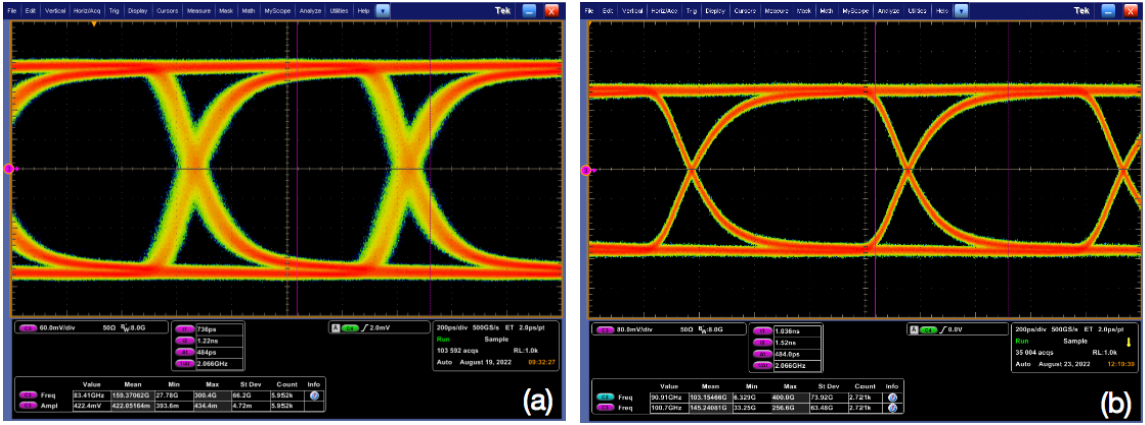

**Figure 5.** Eye diagrams of an uplink channel in the equalizer (a) and retiming (b) modes.

## 5. Conclusion

This paper presents the design and test results of GBCR2 and GBCR3. For the GBCR2, no error is observed in any tested channel in a 400 MeV proton beam. The extrapolated BER to the future HL-LHC application is below $8\times10^{-16}$, significantly less than the specified BER criterion. GBCR3 improves SEE immunity by applying the Triple Modular Redundancy (TMR) technology to all RX channels. The retimed signals from GBCR3 have less total jitter than those from GBCR2 (35 ps versus 79 ps). Each receiver channel of GBCR3 consumes 75% more power than that of GBCR2. The readout full-chain and irradiation tests of GBCR3 will be carried out in the future.

## Acknowledgments

This work is supported by the US-ATLAS phase-2 upgrade program and the Office of High Energy Physics of the U.S. Department of Energy under contract DE-AC02-05CH11231. We are grateful to Ms. Meka E. Francis, Mr. Joel Fulgham, Ms. Kathy Graden, Ms. Caleb McConchie, Ms. Susan McGimpsey, Mr. David Mertz, Mr. Evan Niner, Ms. Kelly Rehr-Scarrio, Ms. Mandy Rominsky, Mr. Eric Schlatter, Ms. Maddie Schoell, Dr. Dong Su, Mr. Michael J Utes, Dr. Zijun Xu, and Mr. Andrew Young, for providing help on the tests.

## References

[1] ATLAS Collaboration, *Technical design report for the ATLAS Inner Tracker Pixel Detector*, CERN-LHCC-2017-021 (2018), ATLAS-TDR-030, CERN, Geneva (2017).

[2] RD53 collaboration, *RD53B users guide*, Tech. Rep., CERN-RD53-PUB-21-001, CERN, Geneva (2020).

[3] N. Guettouche, *The lpGBT production testing system*, 2022 *JINST* 17 C03040.

[4] C. Soos et al., Versatile Link+ Transceiver Production, talk given at *TWEPP 2022 Topical Workshop on Electronics for Particle Physics,* Bergen, Norway, 22 September 2022.

[5] C. Chen et al., *Characterization of a gigabit transceiver for the ATLAS Inner Tracker pixel detector readout upgrade*, 2020 *JINST* 15(03) T03005.

[6] W. Zhang et al., *Characterization and quality control test of a gigabit cable receiver ASIC (GBCR2) for the ATLAS Inner Tracker Detector upgrade*, 2021 *JINST* 16(08) P08013.

[7] M. Dentan et al., *ATLAS Policy on Radiation Tolerant Electronics*, *Report ATC-TE-QA-0001 v.3 (2000), EDMS 113816, 21 July 2020.*